\documentclass[journal]{IEEEtran}
\usepackage{cite} % Tidies up citation numbers.
\usepackage{url} % Provides better formatting of URLs.
\usepackage[utf8]{inputenc} % Allows Turkish characters.
\usepackage[english]{babel}
\usepackage{booktabs} % Allows the use of \toprule, \midrule and \bottomrule in tables for horizontal lines
\usepackage{graphicx}

\usepackage{multirow}
\usepackage{epsfig}
\usepackage{algorithm}
\usepackage{algpseudocode}
\usepackage{amsmath}
\usepackage{amssymb}
\usepackage{url}
\usepackage{caption}
\usepackage{subcaption}
\DeclareCaptionType{copyrightbox}
\usepackage{float}

\algnotext{EndIf}
\algnotext{EndFor}
\algnotext{EndWhile}

\newtheorem{example}{Example}[section]

\begin{document}
\selectlanguage{english}
%\begin{titlepage}
\title{Boosting Frequent Itemset Mining via Early Stopping Intersections}


\author{Huu Hiep Nguyen \\
\{nguyenhuuhiep2@dtu.edu.vn\} \\
Institute of Research and Development, Duy Tan University\\
P809 7/25 Quang Trung, Danang 550000, Vietnam\\
}
%\date{19/08/18}

\maketitle
%\tableofcontents
%\listoffigures
%\listoftables
%\end{titlepage}

\IEEEpeerreviewmaketitle
\begin{abstract}
Mining frequent itemsets from a transaction database has emerged as a fundamental problem in data mining and committed itself as a building block for many pattern mining tasks. In this paper, we present a general technique to reduce support checking time in existing depth-first search generate-and-test schemes such as Eclat/dEclat and PrePost+. Our technique allows infrequent candidate itemsets to be detected early. The technique is based on an early-stopping criterion and is general enough to be applicable in many frequent itemset mining algorithms. We have applied the technique to two TID-list based schemes (Eclat/dEclat) and one N-list based scheme (PrePost+). Our technique has been tested over a variety of datasets and confirmed its effectiveness in runtime reduction.
%\keywords{Frequent itemset mining, early-stopping intersections, Eclat, dEclat, PrePost+}
\end{abstract}

\section{Introduction}
\label{sec:intro}
First proposed by Agrawal et al. \cite{agrawal1994fast}, frequent itemset mining has become a popular data mining technique and has been studied extensively by many researchers. It has played an essential role in many important data mining tasks such as mining association rules \cite{toivonen1996sampling}, sequential patterns \cite{srikant1996mining,fournier2017survey}, correlations \cite{lee2003comine}, episodes \cite{wu2013mining}, classification \cite{nguyen2012classification}, clustering \cite{agrawal1998automatic} and so on. Although lots of algorithms have been proposed, how to improve the efficiency of itemset mining algorithms is still one of several key research problems to be solved. 

Recently, Deng et al. \cite{deng2012new} proposed PrePost and its enhanced version PrePost+ \cite{deng2015prepost+} for mining frequent itemsets. Both of them employ a novel data structure named N-list to represent itemsets and adopt single path property of N-list to directly discover frequent itemsets without generating candidate itemsets in some cases. The experiments in \cite{deng2012new,deng2015prepost+} show that PrePost/PrePost+ run faster than some state-of-the-art mining algorithms including FP-growth \cite{han2000mining} and FP-growth* \cite{grahne2005fast}. By investigating PrePost+, we found that support checking time for candidate itemsets can be reduced largely if we can stop early the N-list intersection for infrequent candidate itemsets. The same idea holds for other schemes that propose and test potential children itemsets by intersecting lists held in parent itemsets. Two such schemes are Eclat \cite{zaki1997new} which uses transaction ID lists (TID-lists) and dEclat \cite{zaki2003fast} which uses Diffsets.

In this work, we further improve Eclat/dEclat and PrePost+ by proposing a simple yet effective technique to stop early the support checking of \textit{infrequent} candidate itemsets in depth-first search. Given an infrequent candidate itemset, the \textit{Early Stopping} technique accumulates the evidence of infrequency and decides early if the candidate's support is undoubtedly less than the minimum support, so further checking steps are redundant and dropped. The runtime reduction is always guaranteed, especially on datasets with high ratio between the number of  candidates and the number of frequent itemsets.

In the next subsection, we review the mainstream of frequent itemset mining.

\subsection{Related Work}
Itemset mining is an important problem of data mining with many variations such as frequent itemset mining \cite{agrawal1994fast,zaki1997new,han2000mining}, frequent closed/maximal itemset mining \cite{zaki2002charm,zaki2005efficient,grahne2005fast}, frequent weighted itemset mining \cite{vo2013new}, erasable itemset mining \cite{deng2009mining} and so on. However, frequent itemset mining is still the most popular as it plays an important role in association rule mining \cite{agrawal1994fast}, sequential mining \cite{fournier2017survey}, classification \cite{nguyen2012classification}. There have been a large number of algorithms which effectively mine frequent itemsets. We may divide them into three main categories:
\begin{itemize}
\item \textbf{Candidate generate-and-test strategy}: Methods in this category use a level-wise (breadth-first-search) approach for mining frequent itemsets. First, they enumerate frequent 1-itemsets which are then used to propose candidate 2-itemsets, and so on until no more candidates can be generated. Apriori \cite{agrawal1994fast} is a seminal work in this line of research.

\item \textbf{Divide-and-conquer strategy}: Methods using this strategy compress the dataset into a summary structure (e.g., FP-Tree, H-struct) and mine frequent itemsets from this structure by using a divide-and-conquer strategy. They do not propose any candidate itemsets. Instead, frequent itemsets are discovered recursively in sub-databases according to the patterns found. FP-Growth \cite{han2000mining}, FP-Growth* \cite{grahne2005fast} and H-Mine\cite{pei2001h} are representative algorithms in this category. All of them run depth-first search.

\item \textbf{Hybrid strategy}: Methods in this category use vertical data formats to summarize the database and mine frequent itemsets by using the generate-and-test strategy. However, the generate-and-test strategy is realized in depth-first manner. TID-list based methods Eclat \cite{zaki1997new}, dEclat \cite{zaki2005efficient}, and N-list-based methods PrePost/PrePost+ \cite{deng2012new,deng2015prepost+} are some typical examples.
\end{itemize}

\subsection{Contributions and Paper Structure}
In this study, we have made the following contributions
\begin{itemize}
\item We point out a common characteristic of depth-first search mining schemes that generate and test candidate itemsets by list intersection.
\item We propose a general and effective Early-Stopping technique for improving list intersection in Eclat, dEclat and PrePost+. The technique always guarantees that the number of comparisons is reduced, leading to runtime cut-down in most of the cases.
\item We have tested the technique over a wide range of datasets and found the cases in which Early-Stopping improves the existing schemes most.
\end{itemize}

The paper is structured as follows. We review the key concepts of frequent itemset mining and the depth-first-search technique in the next section. Our proposed technique will be presented in Sections \ref{sec:eclat} (for Eclat/dEclat) and \ref{sec:prepost} (for PrePost+) followed by the evaluation in Section \ref{sec:exp}. Finally, we conclude the paper and propose future work in Section \ref{sec:conclusion}.

% %
\section{Background}
\label{sec:background}
In this section, we review basic concepts of frequent pattern mining and describe a transaction database as running example. The Early Stopping technique is clarified in the next two sections. 

\subsection{Frequent Itemsets}
We assume a dataset $DB$ consists of $n$ transactions such that each transaction contains a number of items belonging to $I$ where $I=\{i_1, i_2, ..., i_m\}$ is the set of all items in $DB$.

The support of an itemset $X \subseteq I$, denoted by $\rho(X)$, is the number of transactions in $DB$ which contain all the items in $X$. An itemset $X$ is a \textit{frequent itemset} if $\rho(X) \geq minSup $ , where $minSup$ is a given threshold. Note that a frequent itemset with $k$ elements is called a frequent k-itemset, and $F_1$ is the set of frequent 1-itemsets sorted in frequency ascending or descending order.

Table \ref{tab:example} shows a $DB$ of 10 transactions with $I=\{a,b,c,d,e\}$. The $minSup$ is fixed to 3, i.e., itemsets with frequency at least 3 will be output, e.g., $\{a,c\}$ with frequency 4 as it appears in the transactions 3,4,6 and 8. In PrePost+ \cite{deng2015prepost+}, the items are sorted in decreasing frequency as $\{a,c,e,d,b\}$ for PPC-tree because their frequencies are 7,7,7,6, and 3 respectively (see the third column of Table \ref{tab:example}). In the search tree of Eclat/dEclat, the items are sorted in increasing frequency as $\{b,d,a,c,e\}$. These choices of sorting order make the number of candidates as small as possible.

\begin{table}[h!]
\centering
\caption{An example transaction dataset} \label{tab:example}
\begin{tabular}{|l|l|l|}
\hline
Transaction & Items & Reordering in PrePost+ \cite{deng2015prepost+}\\
\hline
1 & a, d, e & a, e, d \\
\hline
2 & b, c, d & c, d, b \\
\hline
3 & a, c, e & a, c, e \\
\hline
4 & a, c, d, e & a, c, e, d \\
\hline
5 & a, e & a, e \\
\hline
6 & a, c, d & a, c, d \\
\hline
7 & b, c & c, b \\
\hline
8 & a, c, d, e & a, c, e, d \\
\hline
9 & b, c, e & c, e, b \\
\hline
10 & a, d, e & a, e, d \\
\hline
\end{tabular}
\setlength{\abovecaptionskip}{-5pt}
\setlength{\belowcaptionskip}{-5pt}
\end{table}	

\subsection{Downward Closure Property and Depth-First-Search}
\begin{figure}
\centering
\includegraphics[height=2.0in]{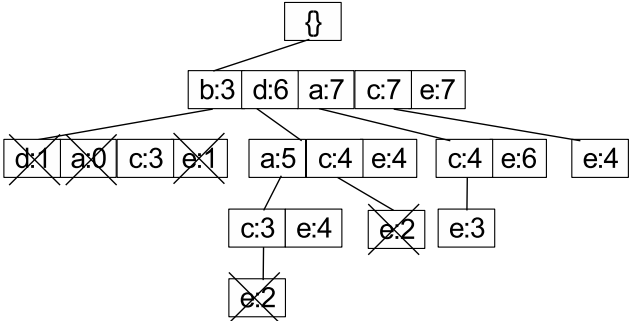}
\setlength{\abovecaptionskip}{-10pt}
\caption{Depth-first-search in Eclat/dEclat. The support is shown after each node's name.}
\vspace{-1.0em}
\label{fig:eclat}
\end{figure}

\textit{Downward closure} (or \textit{anti-monotone}) property \cite{agrawal1994fast}: 
\begin{equation}
\forall X: \forall Y \supseteq X, \rho(Y) \leq \rho(X)
\end{equation}
That means if an itemset is extended, its support cannot increase. In other words, no superset of an infrequent itemset can be frequent. This fact suggests that we can start the search from small itemsets to larger ones. In the search process, if we know that an itemset $X$ is infrequent, we will no longer extend its branch \cite{agrawal1994fast}. In the search tree (Fig. \ref{fig:eclat}), the path from the root to a node represents an itemset under consideration with its support, e.g., itemset $dac$ has support 3.

In depth-first-search schemes like Eclat \cite{zaki1997new}, dEclat \cite{zaki2003fast} and PrePost+ \cite{deng2015prepost+}, the search tree is expanded and visited in depth-first manner. For instance, the order of 15 found frequent itemsets in Fig. \ref{fig:eclat} is: $b, bc, d, da, dac, dae, dc, de, a, ac, ace, ae, c, ce, e$.

In the next sections, we present the Early Stopping technique for Eclat/dEclat and PrePost+ respectively.

% %
\section{Eclat/dEclat With Early Stopping}
\label{sec:eclat}
\subsection{Eclat}
In \cite{zaki1997new}, Zaki et al. proposed Eclat, a depth-first-search technique for frequent itemset mining. Its basic idea is based on \textit{downward closure} as in Apriori but the search is depth-first, not level-wise. 

Eclat uses \textit{vertical format} to represent the database in which each itemset has its own list of transaction ids (TID-list). In Table \ref{tab:vertical}, TID-lists of each item (1-itemset) is a sorted list of transactions containing the item. The TID-list of an itemset $X$ is denoted $T(X)$. We need to read the transaction database once to build the TID-lists of all items. 

We then explore frequent 2-itemsets by intersecting the TID-lists of 1-itemsets. For example, $T(ac) = T(a) \cap T(c) = \{3,4,6,8\}$, so $\rho(ac)=4$. In general, k-itemset $Pxy$ is proposed and tested by intersecting the TID-lists of two (k-1)-itemsets $Px$ and $Py$ (which are both frequent, of course). For example, we have $T(da) = \{1,4,6,8,10\}$ and $T(dc)=\{2,4,6,8\}$, so $T(dac) = T(da) \cap T(dc) = \{4,6,8\}$ and $\rho(dac) = 3$ (frequent).

\begin{table}[h!]
\centering
\caption{Vertical format} \label{tab:vertical}
\begin{tabular}{|l|l|l|l|l|}
\hline
b & d & a & c & e \\
\hline
2 & 1 & 1 & 2 & 1 \\
7 & 2 & 3 & 3 & 3 \\
9 & 4 & 4 & 4 & 4 \\
 &  6 & 5 & 6 & 5\\
 &  8 & 6 & 7 & 8  \\
 & 10 & 8 & 8 & 9\\
 &    & 10 & 9 & 10\\
\hline
\end{tabular}
\setlength{\abovecaptionskip}{-5pt}
\setlength{\belowcaptionskip}{-5pt}
\end{table}

\subsubsection{Early Stopping for Eclat}
Main steps of Eclat are depicted in Algorithm \ref{algo:eclat}. It starts with the creation of TID-list $T(x)$ for each frequent 1-itemset $x$ (Line 2). The depth-first search is delegated to the recursive function TRAVERSE (Lines 8-17). The main step in TRAVERSE is to propose a candidate $Pxy$ (Line 11) and to check its support against $minSup$ (Line 12). 

Looking at the INTERSECT function (Lines 18-29), we found that its runtime is $O(|U|+|V|)$. If $Pxy$ is frequent, we stop only when the condition in Line 20 is violated. However, if $Pxy$ is infrequent, we can stop the intersection early. 

The basic idea is to keep track of \textit{skipped} TIDs in $U$ (called $s_U$) and $V$ (called $s_V$) (see Lines 37 and 41 in the function INTERSECT\_ES). If the number of items that can be matched in $U$ (i.e., $|U| - s_U$) or in $V$ (i.e., $|V| - s_V$) is less than $minSup$, we will surely know that the intersection between $U$ and $V$ is less than $minSup$, resulting in an infrequent candidate itemset. Simply replacing INTERSECT with INTERSECT\_ES helps to reduce the number of comparisons, hence incurring less time to run Eclat.

\begin{example}
\label{ex:tid-intersect-es}
With $T(b)=\{2,7,9\}$ and $T(d)=\{1,2,4,6,8,10\}$, INTERSECT($T(b),T(d)$) stops at $i=4, j=6$ and returns $\{2\}$ while INTERSECT\_ES($T(b),T(d)$) stops at $i=3, j=5$ with $s_U=1, s_V=3$, telling us that $|U|-s_U = 3-1 < minSup$.
\hfill $\Box$
\end{example}

\begin{algorithm}
\caption{Eclat \cite{zaki1997new}} 
\label{algo:eclat}  
\begin{algorithmic}[1]
\Require $DB$ : database with $n$ transactions. $minSup$.
\Ensure $F$, the set of all frequent itemsets
\Procedure{Eclat}{} 	%\Comment{The g.c.d. of a and b}
	\State Scan DB to get $T(x)$ for each frequent item $x$.
	\State $F_1 = F_1 \cup \{T(x)\}$
	\State $F = F \cup \{x | T(x) \in F_1\}$
	\State TRAVERSE($F_1$)
	\State \Return $F$.
\EndProcedure%
\Function{Traverse}{$F_k$}			\Comment{depth-first-search}
	\State $F_{k+1} = \emptyset$
	\For {$T(Px), T(Py) \in F_k$, $x < y$}
		\State $T(Pxy) = $ INTERSECT($T(Px),T(Py)$)
		\If{$|T(Pxy)| \geq minSup$}
			\State $F_{k+1} = F_{k+1} \cup \{T(Pxy)\}$
			\State $F = F \cup \{Pxy\}$
		\EndIf
	\EndFor
	\If{$F_{k+1} != \emptyset$}
		\State TRAVERSE($F_{k+1}$)
	\EndIf
\EndFunction
\Function{Intersect}{$U, V$}
	\State $Z = \emptyset$, $i=1, j=1$
	\While {$i \leq |U|$ AND $j \leq |V| $}
		\If {$U[i] == V[j]$}
			\State $Z = Z \cup \{U[i]\}$
			\State $i++$; $j++$
		\ElsIf {$U[i] < V[j]$}
			\State $i++$
		\Else
			\State $j++$
		\EndIf
	\EndWhile
	\State \Return $Z$.
\EndFunction
\Function{Intersect\_ES}{$U, V$}		\Comment{early-stopping}
	\State $Z = \emptyset$, $i=1, j=1$, $s_U=0, s_V=0$
	\While {$i \leq |U|$ AND $j \leq |V| $}
		\If {$U[i] == V[j]$}
			\State $Z = Z \cup \{U[i]\}$
			\State $i++$; $j++$
		\ElsIf {$U[i] < V[j]$}
			\State $i++$, $s_U++$
			\If{$|U| - s_U < minSup$}
				\State \textbf{break}
			\EndIf
		\Else
			\State $j++$, $s_V++$
			\If{$|V| - s_V < minSup$}
				\State \textbf{break}
			\EndIf
		\EndIf
	\EndWhile
	\State \Return $Z$.
\EndFunction
\end{algorithmic}
\end{algorithm}

\subsection{dEclat}
To reduce memory consumption, Zaki et al. \cite{zaki2003fast} proposed a novel vertical data representation called \textit{Diffset} which only stores differences in the TID-list of a candidate itemset from its generating frequent parents.

From a pair of nodes $Px$,$Py$ having the same prefix $P$ in the search tree, the authors of \cite{zaki2003fast} show that the diffset $D(Pxy) = D(Py) - D(Px)$ and $\rho(Pxy) = \rho(Px) - |D(Pxy)|$, i.e., we can compute the support of an itemset using its parent's support and its own diffset. The diffsets are usually smaller than TID-lists, so the memory consumption is reduced.

\begin{figure}
\centering
\includegraphics[height=2.4in]{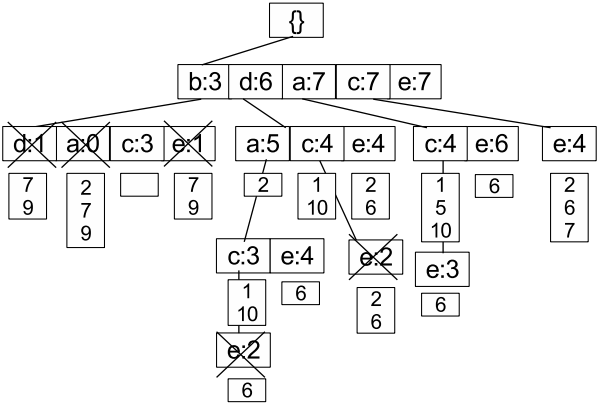}
\setlength{\abovecaptionskip}{-10pt}
\caption{dEclat with diffsets at each node.}
\vspace{-1.0em}
\label{fig:declat}
\end{figure}

Fig. \ref{fig:declat} illustrates such operations on our running example. At the first level, we store $T(x)$ instead of $D(x)$ for all 1-itemsets $x$, especially on sparse databases. At the second level, $D(xy) = T(x) - T(y)$ \cite{zaki2003fast}. For example, $D(bd) = T(b) - T(d) = \{2,7,9\} - \{1,2,4,6,8,10\} = \{7,9\}$, hence, $\rho(bd) = \rho(b) - |D(bd)| = 3 - 2 = 1$ (infrequent).

From the third level, the diffsets are computed directly from parents diffsets, $D(Pxy) = D(Py) - D(Px)$. For example, $D(dac) = D(dc) - D(da) = \{1,10\} - \{2\} = \{1,10\}$, so $\rho(dac) = \rho(da) - |D(dac)| = 5 - 2 = 3$ (frequent).

\subsubsection{Early Stopping for dEclat}
Main steps of dEclat are shown in Algorithm \ref{algo:declat}. Similar to Eclat, it starts with the creation of TID-list $T(x)$ for each frequent 1-itemset $x$ (Line 2). The depth-first search is delegated to the recursive function TRAVERSE (Lines 8-17). The main step in TRAVERSE is to propose a candidate $Pxy$ (Line 11) and to check its support against $minSup$ (Line 12) using the formula $\rho(Pxy) = \rho(Px) - |D(Pxy)|$. 

Looking at the DIFFERENCE function (Lines 18-31), we found that it runs in time $O(|U|+|V|)$. If $Pxy$ is frequent, we stop only when the condition in Line 20 is violated. However, if $Pxy$ is infrequent, we can stop the difference early. 

The basic idea is to check if the support of $Pxy$ is less than $minSup$ after a TID is added to $Z$ (see Lines 40 and 41 in the function DIFFERENCE\_ES). If $\rho_U - |Z| < minSup$, we will surely know that the $Pxy$ is an infrequent candidate itemset. Simply replacing DIFFERENCE with DIFFERENCE\_ES helps to reduce the number of comparisons, hence incurring less runtime of dEclat. Note that compared to the intersection operation which is symmetric in Eclat, the difference operation is asymmetric.

\begin{example}
\label{ex:diffset-intersect-es}
Let $T=\{1,2,3,4,5,6,7,8,9,10\}$ be the set of all TIDs. We have $D(b) = T - T(b) = \{1,3,4,5,6,8,10\}$ and $D(d) = T - T(d) = \{3,5,7,9\}$, DIFFERENCE($D(b),D(d)$) = $D(d) - D(b) = \{7,9\}$ stops at $i=5, j=7$ while DIFFERENCE\_ES($D(b),D(d)$) stops at $i=3, j=6, |Z|=1$, making $\rho(b) - |Z| = 2 < minSup$.
\hfill $\Box$
\end{example}

\begin{algorithm}
\caption{dEclat \cite{zaki2003fast}} 
\label{algo:declat}  
\begin{algorithmic}[1]
\Require $DB$ : database with $n$ transactions. $minSup$.
\Ensure $F$, the set of all frequent itemsets
\Procedure{dEclat}{} 	%\Comment{The g.c.d. of a and b}
	\State Scan DB to get $T(x)$ for each frequent item $x$.
	\State $F_1 = F_1 \cup \{T(x)\}$
	\State $F = F \cup \{x | T(x) \in F_1\}$
	\State TRAVERSE($F_1$)
	\State \Return $F$.
\EndProcedure%
\Function{Traverse}{$F_k$}				\Comment{depth-first-search}
	\State $F_{k+1} = \emptyset$
	\For {$D(Px), D(Py) \in F_k$, $x < y$}
		\State $D(Pxy) = $ DIFFERENCE($D(Py),D(Px)$)
		\If{$\rho(Px) - |D(Pxy)| \geq minSup$}
			\State $F_{k+1} = F_{k+1} \cup \{D(Pxy)\}$
			\State $F = F \cup \{Pxy\}$
		\EndIf
	\EndFor
	\If{$F_{k+1} != \emptyset$}
		\State TRAVERSE($F_{k+1}$)
	\EndIf
\EndFunction
\Function{Difference}{$U, V$}
	\State $Z = \emptyset$, $i=1, j=1$
	\While {$i \leq |U|$ AND $j \leq |V| $}
		\If {$U[i] == V[j]$}
			\State $i++$; $j++$
		\ElsIf {$U[i] < V[j]$}
			\State $Z = Z \cup \{U[i]\}$
			\State $i++$
		\Else
			\State $j++$
		\EndIf
	\EndWhile
	\If{$i \leq |U|$}
		\State $Z = Z \cup \{U[k]|k = i \rightarrow |U|\}$
	\EndIf
	\State \Return $Z$.
\EndFunction
\Function{Difference\_ES}{$U, V, \rho_U$}		\Comment{early-stopping}
	\State $Z = \emptyset$, $i=1, j=1$
	\While {$i \leq |U|$ AND $j \leq |V| $}
		\If {$U[i] == V[j]$}
			\State $i++$; $j++$
		\ElsIf {$U[i] < V[j]$}
			\State $Z = Z \cup \{U[i]\}$
			\State $i++$
			\If{$\rho_U - |Z| < minSup$}
				\State \Return $Z$
			\EndIf
		\Else
			\State $j++$
		\EndIf
	\EndWhile
	\If{$i \leq |U|$}
%		\If{$\rho_U - (|Z| + |U| - i) < minSup$}
%			\State \Return $Z$.
%		\EndIf
		\State $Z = Z \cup \{U[k]|k = i \rightarrow |U|\}$
	\EndIf
	\State \Return $Z$.
\EndFunction
\end{algorithmic}
\end{algorithm}

% %
\section{PrePost+ With Early Stopping}
\label{sec:prepost}
In this section, we summarize main concepts of PrePost+ \cite{deng2015prepost+} such as PPC-Tree, PP-code and N-list. Then we show how to apply Early Stopping to PrePost+.

\subsection{PPC-tree and N-list}
\label{subsec:ppc-tree}

Given a reordered $DB$, PPC-Tree \cite{deng2012new} is a tree structure defined as follows
\begin{itemize}
\item It consists of one root labeled as \textit{null} ($\{\}$), and a set of item prefix subtrees as children of the root.
\item Each node in the item prefix subtree contains five fields: \textit{name}, \textit{frequency}, \textit{childnodes}, \textit{pre}, and \textit{post}. The field \textit{name} registers the item this node represents. The field \textit{frequency} stores the number of transactions containing a path reaching this node. The field \textit{childnodes} registers all children of the node. The field \textit{pre} is the pre-order rank of the node. The field \textit{post} is the post-order rank of the node. For a node, its pre-order is the sequence number of the node when scanning the tree by pre-order traversal and its post-order is the sequence number of the node when scanning the tree by post-order traversal.
\end{itemize}

\begin{figure}
\centering
\includegraphics[height=2.0in]{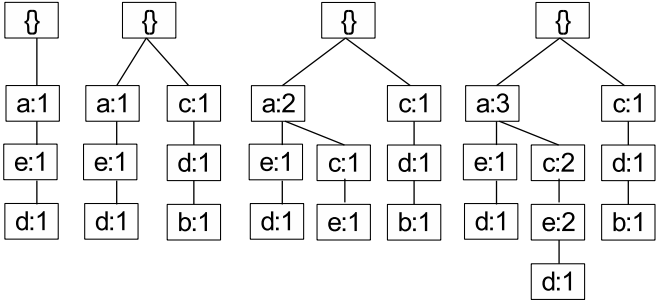}
\setlength{\abovecaptionskip}{-10pt}
\caption{PPC-Tree after inserting first four transactions}
\vspace{-1.0em}
\label{fig:ppc-tree-build}
\end{figure}

\begin{figure}
\centering
\includegraphics[height=2.0in]{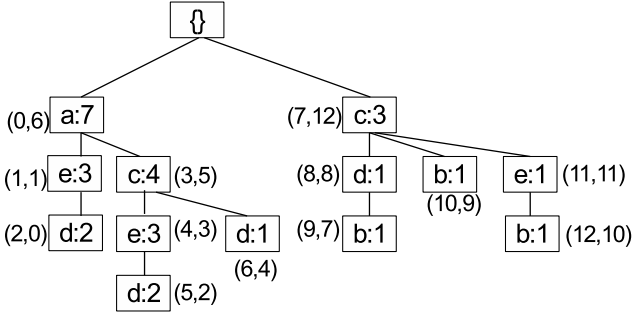}
\setlength{\abovecaptionskip}{-10pt}
\caption{Full PPC-Tree with PP-code of each node}
\vspace{-1.0em}
\label{fig:ppc-tree}
\end{figure}

Fig. \ref{fig:ppc-tree-build} demonstrates how the PPC-Tree is built from the reordered transactions in Table \ref{tab:example}. We start with a null root. Then the first transaction $\{a,e,d\}$ is inserted in the PPC-Tree by creating nodes named $a$, $e$ and $d$ with frequency 1. Similarly, for the second transaction $\{c,d,b\}$, a new child node of the root and two descendent nodes are added. The third and fourth subfigures show the tree after the insertion of $\{a,c,e\}$ and $\{a,c,e,d\}$. The full PPC-Tree is shown in Fig. \ref{fig:ppc-tree}.

The pre-order and post-order ranks are tagged in a pair of numbers next to each node in Fig. \ref{fig:ppc-tree}. 

\textit{PP-code} \cite{deng2012new} of each node $N$ in PPC-Tree is a triple $<N.pre, N.post, N.frequency>$.

\textit{N-list} \cite{deng2012new} of a frequent item $x$, denoted as $NL(x)$ is a sequence of all the PP-codes of nodes $N$ with $N.name=x$ in the PPC-Tree. The PP-codes are arranged in an ascending order of their pre-order ranks. Fig. \ref{fig:n-list-1-item} lists the N-list of all 1-itemsets.

\begin{figure}
\begin{flushleft}
$a \rightarrow <0,6,7>$ \\
$c \rightarrow <3,5,4>,<7,12,3>$ \\
$e \rightarrow <1,1,3>,<4,3,3>,<11,11,1>$ \\
$d \rightarrow <2,0,2>,<5,2,2>,<6,4,1>,<8,8,1>$ \\
$b \rightarrow <9,7,1>,<10,9,1>,<12,10,1>$
\end{flushleft}
\caption{N-list of 1-itemsets}
\label{fig:n-list-1-item}
\end{figure}

Clearly, the support of 1-itemset $x$ is the sum of frequencies of PP-codes in $NL(x)$. For example, $\rho(c) = 4 + 3 = 7, \rho(d) = 2 + 2 + 1 + 1 = 6$.

\textit{N-list of k-itemset} \cite{deng2012new} is defined as follows. Let $xS$ and $yS$ be two (k-1)-itemsets with the same suffix $S$ such that $x$ is before $y$ in frequency ascending ordering. $NL(xS)$ and $NL(yS)$ are two N-lists associated with $xS$ and $yS$, respectively. The N-list associated with $xyS$ is determined as follows (this is performed in \textit{NL\_intersect} function in Algorithm \ref{algo:prepost}):
\begin{enumerate}
\item For each PP-code $X \in NL(xS)$ and $Y \in NL(yS)$, if $Y$ is an ancestor of $X$ in PPC-Tree, the algorithm will add $<Y.pre,Y.post,X.frequency>$ to $NL(xyS)$. Note that the frequency is that of $X$.
\item Traversing $NL(xyS)$ to combine the PP-codes which has the same pre and post values.
\end{enumerate}

\begin{figure}
\centering
\includegraphics[height=2.2in]{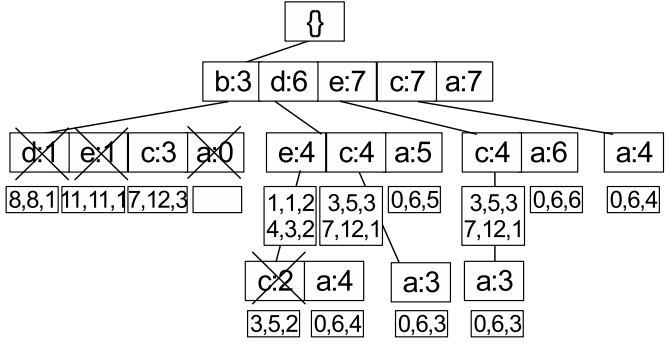}
\setlength{\abovecaptionskip}{-10pt}
\caption{Search tree in PrePost+}
\vspace{-1.0em}
\label{fig:prepost}
\end{figure}

Fig. \ref{fig:prepost} illustrates the PrePost+ search tree for our running example. The items are sorted by descending order in PPC-Tree as $\{a,c,e,d,b\}$ and items are listed the search tree in the reverse order \{b,d,e,c,a\}.

\begin{example}
\label{ex:nl-intersect}
We have $e < c$, $NL(e)=\{<1,1,3>,<4,3,3>,<11,11,1>\}$ and $NL(c)=\{<3,5,4>,<7,12,3>\}$, therefore $NL(ec)=\{<3,5,3>,<7,12,1>\}$ and the support of $ec$ is $\rho(ec) = 3 + 1 = 4$ (see Fig. \ref{fig:prepost}).
\hfill $\Box$
\end{example}

\subsection{PrePost+ Algorithm}
In this section, we briefly recall the PrePost+ algorithm \cite{deng2015prepost+} (see Algorithm \ref{algo:prepost}). PrePost+ starts with the construction of PPC-Tree (Line 1) and computation of NL-list of frequent 1-itemsets (Line 2). Again, the idea of depth-first search in Eclat/dEclat repeats here. Recall that PrePost+ combines itemsets sharing the same suffix (not prefix as in Eclat/dEclat). The recursive function TRAVERSE (Lines 9-18) proposes a candidate $xyS$ (Line 11), computes the intersection between $NL(xS)$ and $NL(yS)$ (Line 12), and checks the support of $xyS$ against $minSup$ (Line 13).

The main steps of \textit{NL\_intersect} are depicted in Lines 19-33 (Algorithm \ref{algo:prepost}). Similar to the function INTERSECTION in Eclat, we maintain two indexes $i$ and $j$ and carry out the intersection from left-to-right. The criteria for the merge (Lines 23,24) are stated in Section \ref{subsec:ppc-tree}, i.e., the $i$-th triple in $U$ is mergeable to the $j$-th triple in $V$ if and only if the former is the ancestor of the latter in the PPC-Tree. 

Again with Example \ref{ex:nl-intersect}, the step-by-step intersection between $NL(e)$ and $NL(c)$ is as follows. $<3,5,4>$ is non-mergeable to $<1,1,3>$ so it is tested against the next $j$, i.e., $<4,3,3>$ where it is mergeable and returns $<3,5,3>$. Then $<3,5,4>$, when compared to $<11,11,1>$, fails at Line 6, so we consider the next $i$, i.e., $<7,12,3>$. Clearly, $<7,12,3>$ is mergeable to $<11,11,1>$, returning $<7,12,1>$. The intersection stops and we get $NL(ec) = \{<3,5,3>, <7,12,1>\}$.

Note that \textit{NL\_intersect} is fixed by the item order, i.e., we only intersect $NL(xS)$ with $NL(yS)$ if $x < y$ in frequency ordering.

\begin{algorithm}                      
\caption{PrePost+ \cite{deng2012new}} 
\label{algo:prepost}                           
\begin{algorithmic}[1]                    
\Require $DB$ : database with $n$ transactions. $minSup$ : minimum support.
\Ensure $F$, the set of all frequent itemsets
\Procedure{PrePost+}{}
	\State Scan DB to obtain $F_1$ and build the PPC-Tree
	\State Scan PPC-tree to generate $NL(x)$
	\State $F_1 = F_1 \cup \{NL(x)\}$
	\State $F = F \cup \{x | NL(x) \in F_1\}$
	\State TRAVERSE($F_1$)
	\State \Return $F$.
\EndProcedure%
\Function{Traverse}{$F_k$}			\Comment{depth-first-search}
	\State $F_{k+1} = \emptyset$
	\For {$NL(xS), NL(yS) \in F_k$, $x < y$}
		\State $NL(xyS) = $ NL\_intersect($NL(xS),NL(yS)$)
		\If{$\rho(xyS) \geq minSup$}
			\State $F_{k+1} = F_{k+1} \cup \{NL(xyS)\}$
			\State $F = F \cup \{xyS\}$
		\EndIf
	\EndFor
	\If{$F_{k+1} != \emptyset$}
		\State TRAVERSE($F_{k+1}$)
	\EndIf
\EndFunction
\Function{NL\_intersect}{$U,V$}
	\State $i=1$, $j=1$
	\State $Z = \emptyset$
	\While {$x_i \in U, i \leq |U| \;\text{AND}\; y_j \in V, j \leq |V|$} 
		\If {$x_i.pre > y_j.pre $}
			\If { $x_i.post < y_j.post$}
				\State add $<y_j.pre, y_j.post, x_i.freq>$ to $Z$
				\State $i++$
			\Else
				\State $j++$
			\EndIf	
		\Else
			\State $i++$	
		\EndIf 
	\EndWhile	
	\State merge elements in $Z$
    \State \Return $Z$.
\EndFunction
\Function{NL\_intersect\_ES}{$U,V,\rho_V$}	\Comment{early-stopping}
	\State $i=1$, $j=1$
	\State $skip = 0$
	\State $Z = \emptyset$
	\While {$x_i \in U, i \leq |U| \;\text{AND}\; y_j \in V, j \leq |V|$} 
		\If {$x_i.pre > y_j.pre $}
			\If { $x_i.post < y_j.post$}
				\State add $<y_j.pre, y_j.post, x_i.freq>$ to $Z$
				\State $i++$
			\Else
				\State $skip = skip + y_i.freq$
				\If {$\rho_V - skip < minSup$}
					\State \Return $\emptyset$
				\EndIf	
				\State $j++$
			\EndIf	
		\Else			
			\State $i++$	
		\EndIf 
	\EndWhile	
	\State merge elements in $Z$
    \State \Return $Z$.
\EndFunction
\end{algorithmic}
\end{algorithm}

\subsection{Early Stopping for PrePost+}
In PrePost+, \textit{NL\_intersect} runs in $O(|U| + |V|)$. To apply Early Stopping technique, we integrate again the size test into the function \textit{NL\_intersect} in order that if the test fails early, we can stop the computation and return an empty $Z$. 

We present this idea in the function \textit{NL\_intersect\_ES} (Lines 34-52). At any triple $j$ of $V$, if it is non-mergeable to the triple $i$ of $U$, we increase $skip$ by $y_j.freq$ (Line 44). If the sum of remaining frequencies $\rho_V - skip$ is less than $minSup$, we stop and return an empty set (Lines 45,46). We demonstrate the effectiveness of \textit{NL\_intersect\_ES} in the next example.

\begin{example}
\label{ex:nl-intersect-es}
Given $NL(b) = \{<9,7,1><10,9,1><12,10,1>\}$ and $NL(d) = \{<2,0,2><5,2,2><6,4,1><8,8,1>\}$, if we call \textit{NL\_intersect(NL(b),NL(d))}, we need to run 5 checks for $(i,j) = (1,1),(1,2),(1,3),(1,4),(2,4)$ in which only pair $(1,4)$ matches, so we get $NL(bd)=\{<8,8,1>\}$. With support 1 (less than $minSup = 3$), $db$ is infrequent.

In calling \textit{NL\_intersect\_ES(NL(b),NL(d))}, we know that $\rho_V = 6, minSup = 3$. After the two (failed) checks $(i,j) = (1,1),(1,2)$, we increase $skip$ to $2 + 2 = 4$, making $\rho_V - skip < minSup$, so we safely conclude that $bd$ is not frequent, omitting the three remaining checks.
\hfill $\Box$
\end{example}

\subsection{Remarks on Apriori, FP-Growth, and Bit-Vector Based Algorithms}
Apriori \cite{agrawal1994fast} is a level-wise (breadth-first search) mining scheme which use \textit{horizontal} format to count the support for candidate $k$-itemsets (i.e., itemsets at level $k$). No list intersection is required in Apriori, so our technique does not apply.

Instead of generating and testing candidate itemsets, FP-Growth \cite{han2000mining} and its derivatives FP-Growth* \cite{grahne2005fast}, H-Mine \cite{pei2001h} recursively project the database into sub-databases using prefix itemsets. Then local frequent patterns are searched to assemble longer global ones. No list intersection is required in FP-Growth/FP-Growth* or H-Mine, so our technique does not apply either.

Bit-vector based algorithm such as VIPER \cite{shenoy2000turbo} applies depth-first search in the same manner as Eclat but uses a compressed bit-vector structure instead. The intersection of decompressed bit-vectors in memory is performed by AND operator. Our technique can be plugged to such algorithms to early determine if the intersection would be less than $minSup$ or not.

\section{Experiments}
\label{sec:exp}
In this section, we evaluate the performance of the proposed Early-Stopping technique applied to Eclat/dEclat and PrePost+ in terms of runtime and number of comparisons. The datasets are described in Sections \ref{subsec:eval-data}. We show the comparison between standard versions and early-stopping versions in Section \ref{subsec:eval-result}. The algorithms are implemented in C++ and run on a desktop PC with $Intel^{\circledR}$ Core i7-6700@ 3.4Ghz, 16GB memory.

\subsection{Experiment Setup}
\label{subsec:eval-data}

We use nine datasets as shown in Table \ref{tab:data-prop}. The datasets were downloaded from FIMI repository (\url{http://fimi.ua.ac.be}) and KONECT repository (\url{http://konect.uni-koblenz.de/networks/}). The columns \#Items, \#Trans, Avg.Length and minSup show the number of items, number of transactions, average transaction length and the range of $minSup$ value respectively.

\textit{T40I10D100K} is a synthetic market-basket dataset from \cite{agrawal1994fast}. It contains 100,000 transactions and 942 items.

\textit{MovieLens-1M} is a bipartite network containing one million movie ratings from \url{http://movielens.umn.edu/}. Movies play the role of items and ratings of each user stand for a transaction.

\textit{Github} is a membership network of the hosting site GitHub. The network is bipartite and contains users (transactions) and projects (items).

\textit{Retail} is anonymous retail market-basket data from an anonymous Belgian retail store.

\textit{Kosarak} contains sequences of click-stream data from a Hungarian news portal.

\textit{Accidents} contains anonymous traffic accident data.

\textit{Chess} is converted from UCI chess dataset. Each transaction is an instance of the chess game and items describe the board and the outcome of the game.

\textit{Connect} is converted from UCI connect-4 dataset. Each transaction is an instance of the game and items describe the board and the outcome of the game.

\textit{Pumsb} dataset contains census data for population and housing. 

We name the Early-Stopping versions as Eclat-ES, dEclat-ES and PrePost+ES. Recall that our technique is easily plugged to any frequent pattern mining schemes that require the intersection operation for itemset support checking.

\begin{table}[!t]
\centering
\caption{Dataset properties} \label{tab:data-prop}
\begin{tabular}{|l|r|r|r|r|}
\hline
\textbf{Dataset} & \textbf{\#Items} & \textbf{\#Trans} & \textbf{Avg.Length} & \textbf{minSup}\\
\hline
T40I10D100K & 942 & 100,000 & 39.6 & 0.002 .. 0.02\\
MovieLens-1M & 3,706 & 6,040 & 165.6 & 0.07 .. 0.1\\
Github & 56,519 & 120,867 & 3.6 & 0.00007 .. 0.0001\\
Retail & 16,470 & 88,162 & 10.3 & 0.00003 .. 0.00006\\
Kosarak & 41,270 & 990,002 & 8.1 & 0.001 .. 0.004\\
Accidents & 468 & 340,183 & 33.8 & 0.1 .. 0.4\\
Chess & 75 & 3,196 & 37.0 & 0.1 .. 0.4 \\
Connect & 129 & 67,557 & 43.0 & 0.1 .. 0.4 \\
Pumsb & 2,088 & 49,046 & 50.5 & 0.1 .. 0.4\\
\hline

\hline
\end{tabular}
\setlength{\abovecaptionskip}{-5pt}
\setlength{\belowcaptionskip}{-5pt}
\end{table}

\subsection{Effectiveness of Early Stopping Technique}
\label{subsec:eval-result}
In this section, we evaluate the effectiveness of Early-Stopping schemes Eclat-ES, dEclat-ES and PrePost+ES. Because the schemes run deterministically, all the reported values except runtime do not change. The runtime is the average result of ten runs.

\subsubsection{Number of Proposed Candidates and Expanded Nodes}

\begin{table*}[!t]
\centering
\caption{Number of proposed candidates and expanded nodes} \label{tab:ratio}
\begin{tabular}{|l|r|r|r|r|r|r|r|r|r|r|r|r|}
\hline
\textbf{Dataset} & \multicolumn{3}{c|}{\textbf{minSup} = $minSup_1$} & \multicolumn{3}{c|}{\textbf{minSup = $minSup_2$}} & \multicolumn{3}{c|}{\textbf{minSup = $minSup_3$}} & \multicolumn{3}{c|}{\textbf{minSup = $minSup_4$}}\\
\hline
				& \#Cands & \#Nodes & Ratio & \#Cands & \#Nodes & Ratio & \#Cands & \#Nodes & Ratio & \#Cands & \#Nodes & Ratio \\
\hline
T40I10D100K & 2.16e+07 & 8.99e+06 & 2.41 & 3.22e+06 & 1.28e+06 & 2.51 & 3.31e+05 & 6.52e+04 & 5.07 & 1.11e+04 & 2.29e+03 & 4.84 \\
MovieLens-1M& 4.24e+07 & 2.29e+07 & 1.85 & 1.00e+07 & 4.86e+06 & 2.06 & 3.12e+06 & 1.36e+06 & 2.30 & 1.16e+06 & 4.55e+05 & 2.55 \\
Github		& 1.11e+08 & 1.94e+07 & 5.71 & 4.99e+07 & 8.30e+06 & 6.02 & 2.65e+07 & 4.07e+06 & 6.51 & 1.00e+07 & 1.32e+06 & 7.60 \\
Retail		& 3.05e+07 & 1.84e+06 & 16.61 & 1.28e+07 & 9.68e+05 & 13.24 & 6.57e+06 & 6.10e+05 & 10.76 & 3.84e+06 & 4.26e+05 & 9.00 \\
Kosarak		& 9.31e+05 & 6.26e+05 & 1.49 & 4.64e+04 & 3.59e+04 & 1.29 & 6.37e+03 & 4.98e+03 & 1.28 & 3.00e+03 & 2.52e+03 & 1.19 \\
Accidents	& 1.01e+07 & 9.96e+06 & 1.02 & 9.08e+05 & 8.87e+05 & 1.02 & 1.55e+05 & 1.50e+05 & 1.03 & 3.40e+04 & 3.25e+04 & 1.04 \\
Chess		& 1.65e+08 & 1.55e+08 & 1.06 & 2.75e+07 & 2.59e+07 & 1.06 & 6.06e+06 & 5.71e+06 & 1.06 & 1.50e+06 & 1.41e+06 & 1.06 \\
Connect		& 1.01e+07 & 8.04e+06 & 1.26 & 1.84e+06 & 1.48e+06 & 1.24 & 5.78e+05 & 4.60e+05 & 1.26 & 2.96e+05 & 2.39e+05 & 1.24 \\
Pumsb		& 7.23e+06 & 5.47e+06 & 1.32 & 3.98e+05 & 2.96e+05 & 1.34 & 4.01e+04 & 2.92e+04 & 1.37 & 5.25e+03 & 3.73e+03 & 1.41 \\
\hline
\end{tabular}
\setlength{\abovecaptionskip}{-5pt}
\setlength{\belowcaptionskip}{-5pt}
\end{table*}

Table \ref{tab:ratio} displays the number of proposed candidates (column \#Cands), expanded nodes (column \#Nodes) and the ratio between them (column Ratio) on the search tree for each dataset and different values of $minSup$. Here $minSup_1$ means the smallest value of $minSup$ for the dataset, $minSup_2$ means the next value and so on (see Table \ref{tab:data-prop}). 

Because Eclat/dEclat and PrePost+ traverse the search tree based on items sorted in increasing frequency, the number of proposed candidates and expanded nodes are the same for all the schemes on a given dataset and $minSup$. As $minSup$ increases, there are less frequent 1-itemsets, so the number of proposed candidates and expanded nodes get smaller too.

We can roughly divide the datasets into two groups by the ratio between the number of candidates and expanded nodes. The first four datasets have the ratio larger than 2 while the remaining have the ratio less than 1.5. As we will see in the following subsections, the ratio suggests different behaviours of mining schemes in both the number of comparisons and runtime.

\begin{figure*}
 	\centering	
         \begin{subfigure}[b]{0.48\textwidth}
                 \centering
                 \epsfig{file=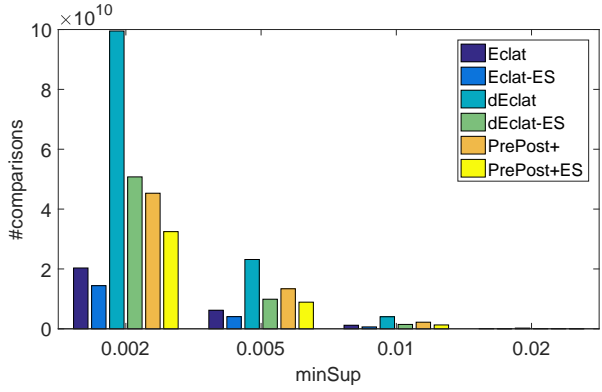, height=2.0in}
%                 \setlength{\abovecaptionskip}{-10pt}
%                 \caption{Soybean}	
                 \label{fig:T40I10D100K-numChecks}
         \end{subfigure}
         \hfill
         \begin{subfigure}[b]{0.48\textwidth}
                 \centering
                 \epsfig{file=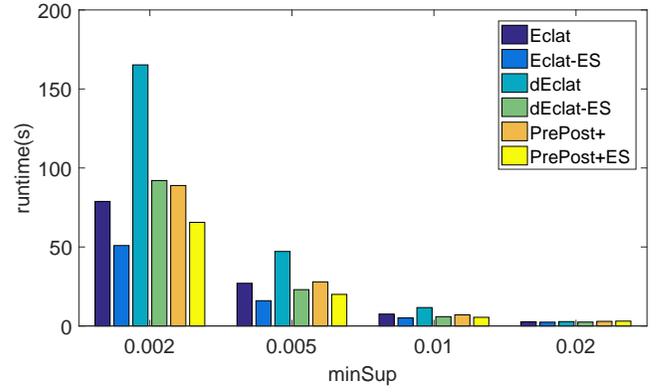, height=2.0in}
%                 \setlength{\abovecaptionskip}{-10pt}
%                 \caption{Mushroom}	
                 \label{fig:T40I10D100K-runtime}
         \end{subfigure}
     \caption{Number of comparisons and runtime for T40I10D100K}
     \label{fig:T40I10D100K}
\end{figure*}

\begin{figure*}
 	\centering	
         \begin{subfigure}[b]{0.48\textwidth}
                 \centering
                 \epsfig{file=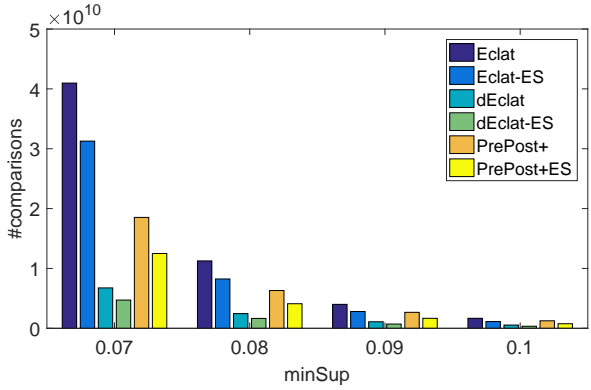, height=2.0in}
%                 \setlength{\abovecaptionskip}{-10pt}
%                 \caption{Soybean}	
                 \label{fig:movielens-1m-numChecks}
         \end{subfigure}
         \hfill
         \begin{subfigure}[b]{0.48\textwidth}
                 \centering
                 \epsfig{file=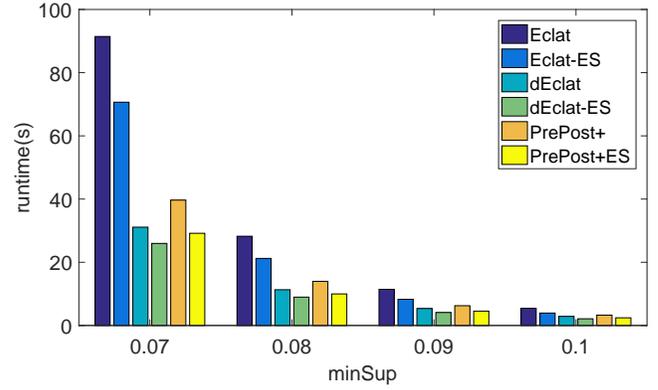, height=2.0in}
%                 \setlength{\abovecaptionskip}{-10pt}
%                 \caption{Mushroom}	
                 \label{fig:movielens-1m-runtime}
         \end{subfigure}
     \caption{Number of comparisons and runtime for MovieLens-1M}
     \label{fig:movielens-1m}
\end{figure*}

\begin{figure*}
 	\centering	
         \begin{subfigure}[b]{0.48\textwidth}
                 \centering
                 \epsfig{file=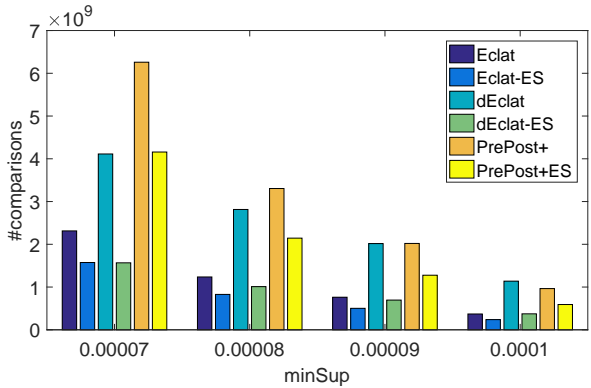, height=2.0in}
%                 \setlength{\abovecaptionskip}{-10pt}
%                 \caption{Soybean}	
                 \label{fig:github-numChecks}
         \end{subfigure}
         \hfill
         \begin{subfigure}[b]{0.48\textwidth}
                 \centering
                 \epsfig{file=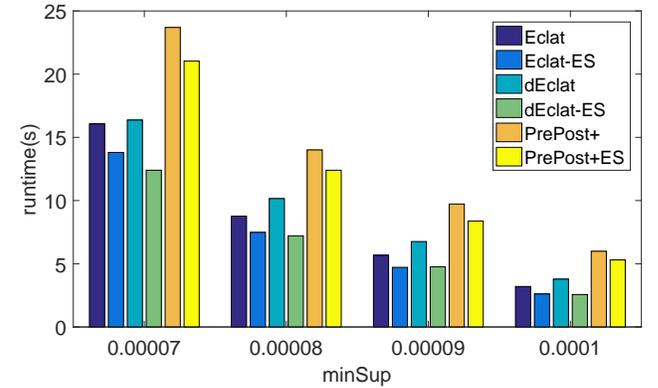, height=2.0in}
%                 \setlength{\abovecaptionskip}{-10pt}
%                 \caption{Mushroom}	
                 \label{fig:github-runtime}
         \end{subfigure}
     \caption{Number of comparisons and runtime for Github}
     \label{fig:github}
\end{figure*}

\begin{figure*}
 	\centering	
         \begin{subfigure}[b]{0.48\textwidth}
                 \centering
                 \epsfig{file=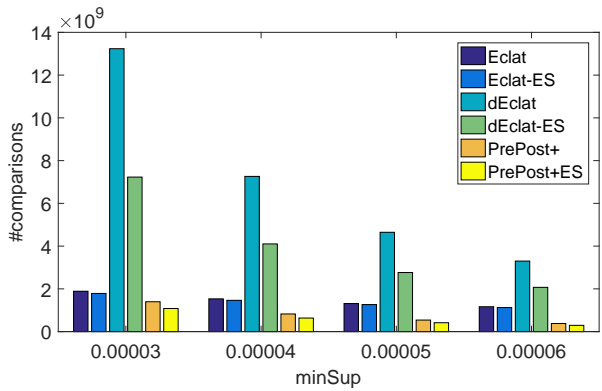, height=2.0in}
%                 \setlength{\abovecaptionskip}{-10pt}
%                 \caption{Soybean}	
                 \label{fig:retail-numChecks}
         \end{subfigure}
         \hfill
         \begin{subfigure}[b]{0.48\textwidth}
                 \centering
                 \epsfig{file=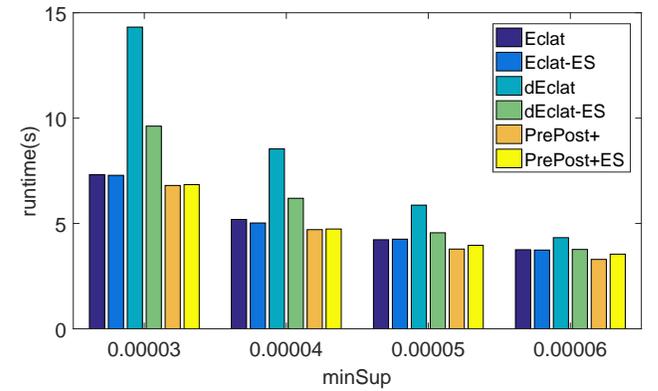, height=2.0in}
%                 \setlength{\abovecaptionskip}{-10pt}
%                 \caption{Mushroom}	
                 \label{fig:retail-runtime}
         \end{subfigure}
     \caption{Number of comparisons and runtime for Retail}
     \label{fig:retail}
\end{figure*}

\begin{figure*}
 	\centering	
         \begin{subfigure}[b]{0.48\textwidth}
                 \centering
                 \epsfig{file=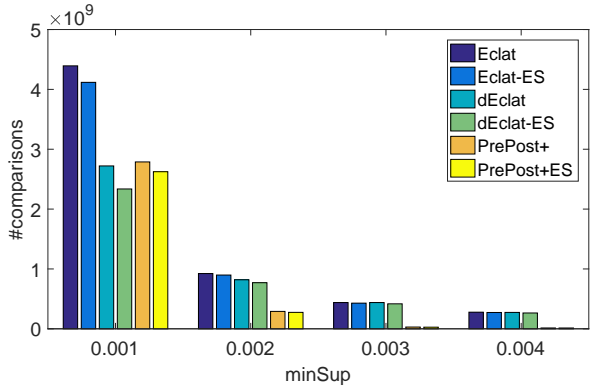, height=2.0in}
%                 \setlength{\abovecaptionskip}{-10pt}
%                 \caption{Soybean}	
                 \label{fig:kosarak-numChecks}
         \end{subfigure}
         \hfill
         \begin{subfigure}[b]{0.48\textwidth}
                 \centering
                 \epsfig{file=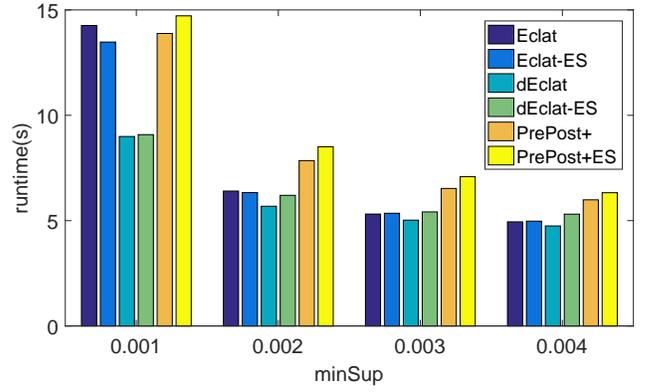, height=2.0in}
%                 \setlength{\abovecaptionskip}{-10pt}
%                 \caption{Mushroom}	
                 \label{fig:kosarak-runtime}
         \end{subfigure}
     \caption{Number of comparisons and runtime for Kosarak}
     \label{fig:kosarak}
\end{figure*}

\begin{figure*}
 	\centering	
         \begin{subfigure}[b]{0.48\textwidth}
                 \centering
                 \epsfig{file=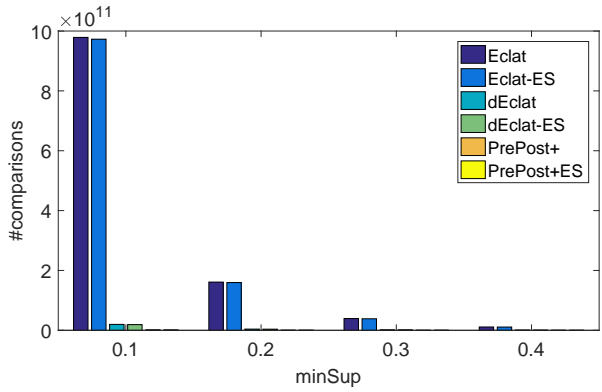, height=2.0in}
%                 \setlength{\abovecaptionskip}{-10pt}
%                 \caption{Soybean}	
                 \label{fig:accidents-numChecks}
         \end{subfigure}
         \hfill
         \begin{subfigure}[b]{0.48\textwidth}
                 \centering
                 \epsfig{file=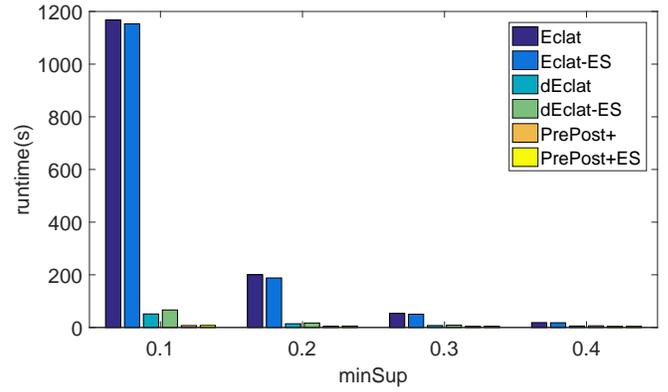, height=2.0in}
%                 \setlength{\abovecaptionskip}{-10pt}
%                 \caption{Mushroom}	
                 \label{fig:accidents-runtime}
         \end{subfigure}
     \caption{Number of comparisons and runtime for Accidents}
     \label{fig:accidents}
\end{figure*}

\begin{figure*}
 	\centering	
         \begin{subfigure}[b]{0.48\textwidth}
                 \centering
                 \epsfig{file=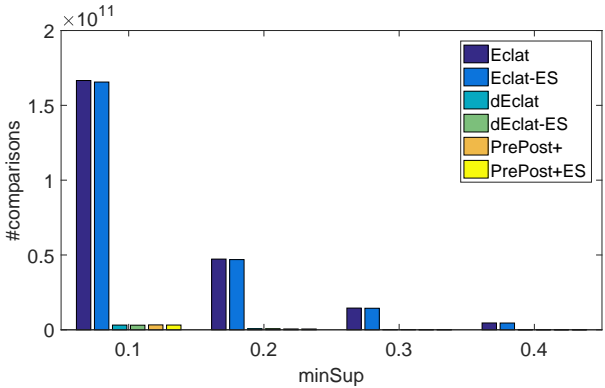, height=2.0in}
%                 \setlength{\abovecaptionskip}{-10pt}
%                 \caption{Soybean}	
                 \label{fig:chess-numChecks}
         \end{subfigure}
         \hfill
         \begin{subfigure}[b]{0.48\textwidth}
                 \centering
                 \epsfig{file=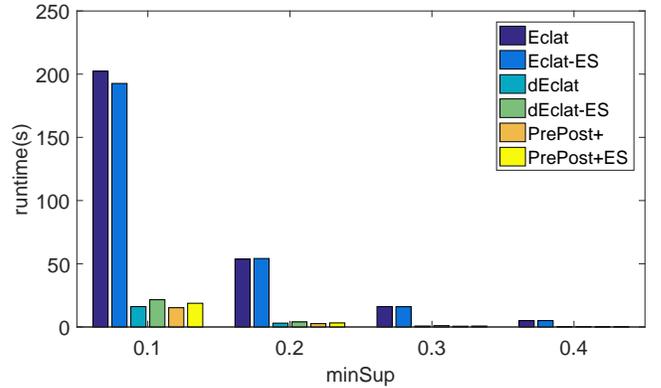, height=2.0in}
%                 \setlength{\abovecaptionskip}{-10pt}
%                 \caption{Mushroom}	
                 \label{fig:chess-runtime}
         \end{subfigure}
     \caption{Number of comparisons and runtime for Chess}
     \label{fig:chess}
\end{figure*}

\begin{figure*}
 	\centering	
         \begin{subfigure}[b]{0.48\textwidth}
                 \centering
                 \epsfig{file=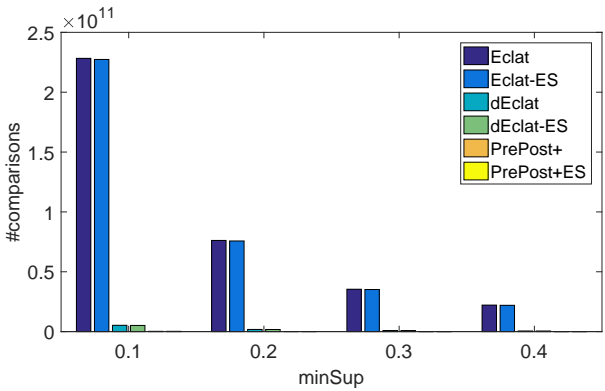, height=2.0in}
%                 \setlength{\abovecaptionskip}{-10pt}
%                 \caption{Soybean}	
                 \label{fig:connect-numChecks}
         \end{subfigure}
         \hfill
         \begin{subfigure}[b]{0.48\textwidth}
                 \centering
                 \epsfig{file=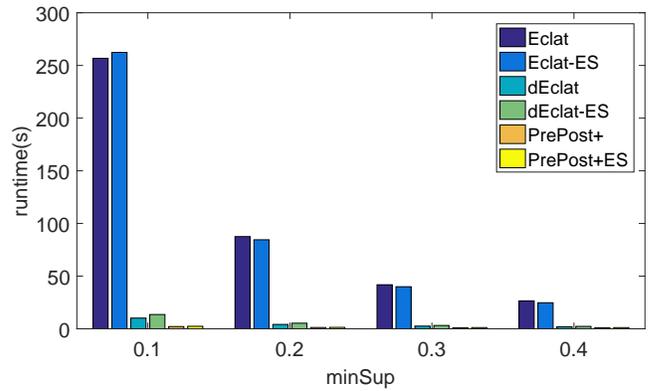, height=2.0in}
%                 \setlength{\abovecaptionskip}{-10pt}
%                 \caption{Mushroom}	
                 \label{fig:connect-runtime}
         \end{subfigure}
     \caption{Number of comparisons and runtime for Connect}
     \label{fig:connect}
\end{figure*}

\begin{figure*}
 	\centering	
         \begin{subfigure}[b]{0.48\textwidth}
                 \centering
                 \epsfig{file=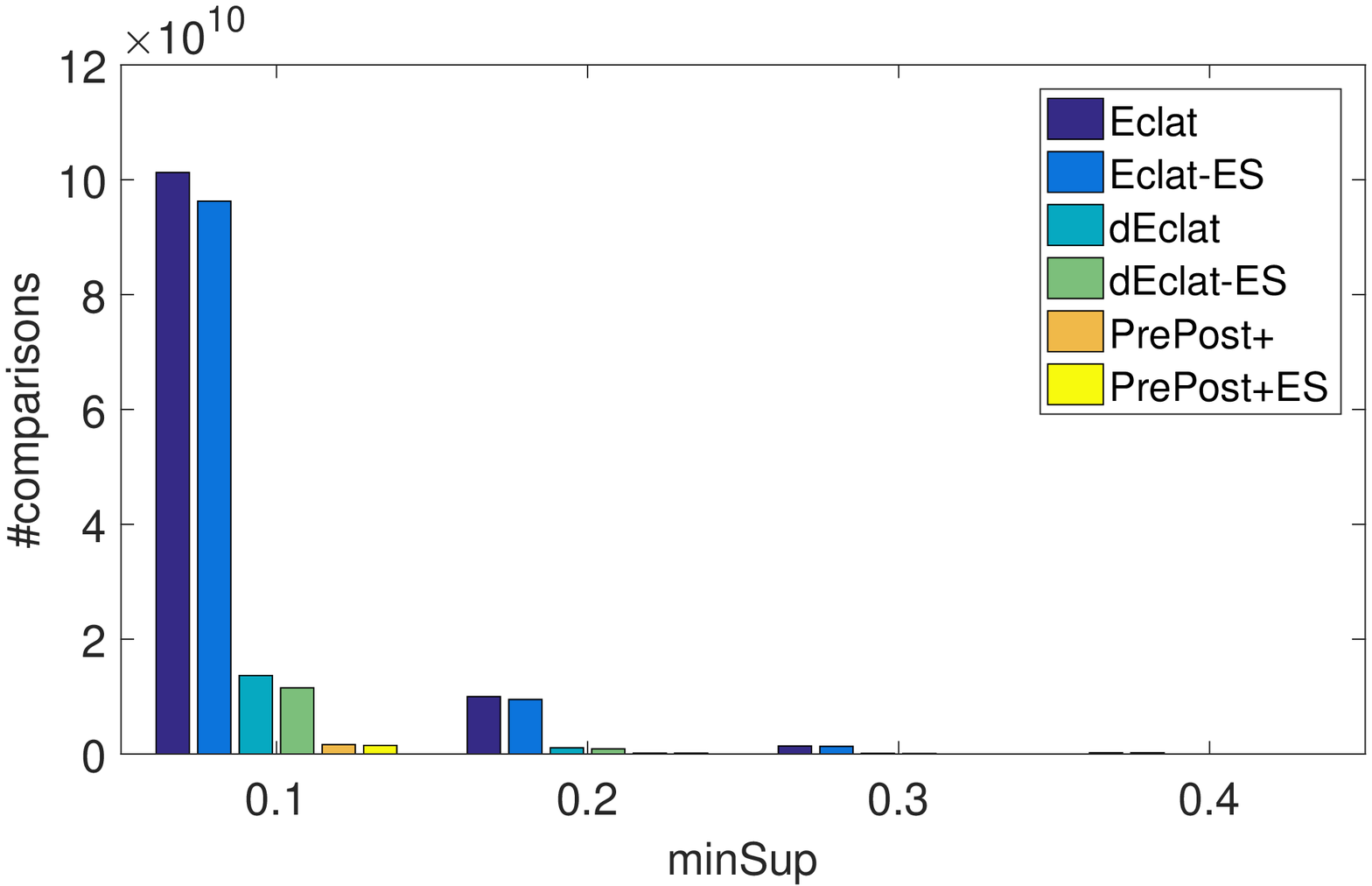, height=2.0in}
%                 \setlength{\abovecaptionskip}{-10pt}
%                 \caption{Soybean}	
                 \label{fig:pumsb-numChecks}
         \end{subfigure}
         \hfill
         \begin{subfigure}[b]{0.48\textwidth}
                 \centering
                 \epsfig{file=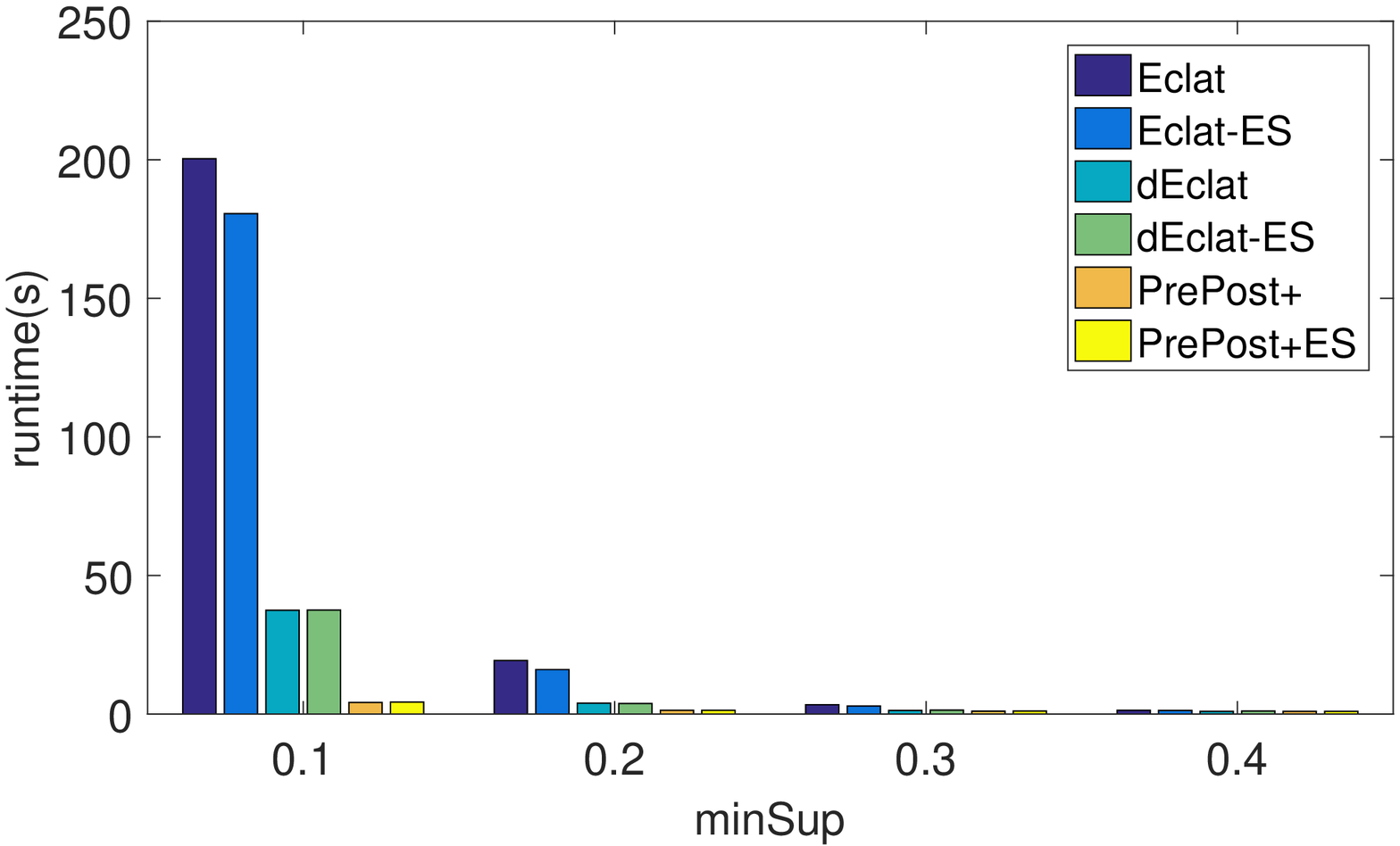, height=2.0in}
%                 \setlength{\abovecaptionskip}{-10pt}
%                 \caption{Mushroom}	
                 \label{fig:pumsb-runtime}
         \end{subfigure}
     \caption{Number of comparisons and runtime for Pumsb}
     \label{fig:pumsb}
\end{figure*}

\subsubsection{Number of Comparisons}
Figures \ref{fig:T40I10D100K} to \ref{fig:pumsb} compare six schemes over nine datasets for different values of $minSup$. In each figure, we report the number of comparisons performed in intersection functions on the left and the total runtime (in second) on the right.

First, the Early-Stopping schemes effectively reduce the number of comparisons between pairs of TID-lists (Eclat), Diffsets (dEclat) or N-lists (PrePost+) in all cases. The reduction varies among datasets and mining schemes. 

For Eclat-ES, the number of comparisons is cut down considerably in the first three datasets T40I10D100K, MovieLens-1M and Github. The reduction is clear cut for small values of $minSup$ and slightly decreases when $minSup$ becomes larger. 

dEclat-ES and PrePost+ES confirm similar effective reduction on T40I10D100K, MovieLens-1M, Github, Retail and Kosarak. 

Finally, we observe that the reduction of comparison operations in Accidents, Chess, Connect and Pumsb is almost negligible. This result can be explained by the ratio column of Table \ref{tab:ratio}. These datasets also exhibit large discrepancies between Eclat and dEclat/PrePost+, confirming that TID-list is much less efficient on these kinds of transaction data.

\subsubsection{Runtime}
The reduction in the number of comparisons naturally translates into the reduction of runtime (see the right plots of Figures \ref{fig:T40I10D100K} to \ref{fig:pumsb}). The clear effect is observed in the four datasets and in Eclat-ES for the remaining five datasets.

Note that the reduction in runtime must take into account the offset caused by the Early-Stopping checks (i.e., Lines 38 and 42 in Algorithm \ref{algo:eclat}, Line 40 in Algorithm \ref{algo:declat} and Line 45 in Algorithm \ref{algo:prepost}). If the candidate itemset is frequent, such checks make Early-Stopping intersection functions incur a small overhead compared to the standard counterparts. For datasets whose number of comparisons is not much saved, the runtime reduction is not guaranteed. This fact is clearly observed in several cases, especially for dEclat-ES and PrePost+ES on Kosarak.

\subsubsection{Other Remarks}
As TID-lists (in Eclat) and Diffsets (in dEclat) are two complementary structures, we observe an interesting tendency: the high number of comparisons or runtime in one scheme implies the low corresponding values in the other. 

All the enhanced versions have the same memory consumption as the original schemes. This fact is straightforward because the memory requirement to maintain the support structures like TID-lists, Diffsets and N-lists is unchanged. The number of proposed candidates does not change either.

% %
\section{Conclusion}
\label{sec:conclusion}
We have presented a simple yet effective Early-Stopping technique to accelerate some existing depth-first search itemset mining algorithms that use the generate-and-test strategy. Our technique is based on an early-stopping criterion for list intersection. We have applied the technique to TID-list in Eclat, diffsets in dEclat and N-list in PrePost+. The number of comparisons in the enhanced versions is always less than that in the original algorithms, leading to runtime cut-down in most of the cases. We have evaluated the Early-Stopping schemes over nine datasets. The results confirm the effectiveness of our improvement and suggest what kind of transaction data will benefit most.


\section*{Conflicts of Interest}
The author(s) declare(s) that there is no conflicts of interest regarding the publication of this paper.


\bibliographystyle{abbrv}
\bibliography{freqpattern}  % diffpriv.bib is the name of the Bibliography in this case

% This is a hand-made bibliography. If you want to use a BibTeX file, you're on your own ;-)

\end{document}